\documentclass[aps,prc,reprint,groupedaddress,showpacs]{revtex4-1}

\usepackage{graphicx}
\usepackage{subfigure}
\usepackage{dcolumn}
\usepackage{bm}
\usepackage{morefloats}

\begin{document}
\title{Multi-Nucleon Exchange in Quasi-Fission Reactions}
\author{S. Ayik$^{1}$}\email{ayik@tntech.edu}
\author{B. Yilmaz$^{2}$}\email{bulent.yilmaz@science.ankara.edu.tr}
\author{O. Yilmaz$^{3}$}\email{oyilmaz@metu.edu.tr}
\affiliation{
$^{1}$Physics Department, Tennessee Technological University, Cookeville, TN 38505, USA \\
$^{2}$Physics Department, Faculty of Sciences, Ankara University, 06100 Ankara, Turkey \\
$^{3}$Physics Department, Middle East Technical University, 06800 Ankara, Turkey }
\date{\today}

\begin{abstract}
Nucleon exchange mechanism is investigated in the central collisions of ${}^{40}$Ca + ${}^{238}$U and ${}^{48}$Ca + ${}^{238}$U systems near the quasi-fission regime in the framework of the Stochastic Mean-Field (SMF) approach. Sufficiently below the fusion barrier, di-nuclear structure in the collisions is maintained to a large extend. Consequently, it is possible to describe nucleon exchange as a diffusion process familiar from deep-inelastic collisions. Diffusion coefficients for proton and neutron exchange are determined from the microscopic basis of the SMF approach in the semi-classical framework. Calculations show that after a fast charge equilibration the system drifts toward symmetry over a very long interaction time. Large dispersions of proton and neutron distributions of the produced fragments indicate that diffusion mechanism may help to populate heavy trans-uranium elements near the quasi-fission regime in these collisions.
\end{abstract}

\maketitle

\section{Introduction}

Much experimental effort has been spent to synthesize super-heavy elements by fusion mechanism of heavy systems \cite{R1,R2,R3,R4,R5}. It is crucial to choose the right projectile-target combinations in order to have the largest probability for forming a compound nucleus that leads to production of the desired super-heavy element. Among different possibilities, formation of compound nucleus in collisions with deformed actinide targets with neutron rich projectile, so called hot fusion, provides a suitable choice for synthesizing of these elements \cite{R1,R2,R3}.  However, in heavy systems compound nucleus formation is severely inhibited by the quasi-fission mechanism \cite{R6,R7}. In quasi-fission, the colliding ions attach together for a long time, but separate without going through compound nucleus formation. During the long contact times many nucleon exchanges take place between projectile and target nuclei. A number of models are developed for a description of the reaction mechanism in the quasi-fission process \cite{R8,R9,R10}. In a recent work, the quasi-fission mechanism in ${}^{40}$Ca + ${}^{238}$U and ${}^{48}$Ca + ${}^{238}$U collisions was investigated in the mean-field approach of the time-dependent Hartree-Fock theory by Oberacker et al. \cite{R11}. The calculations carried out at bombarding energies around the fusion barrier exhibit an important difference between the collisions ${}^{40}$Ca + ${}^{238}$U and ${}^{48}$Ca + ${}^{238}$U. In the collisions with neutron rich isotope of calcium, for increasing bombarding energy quasi-fission mechanism diminishes. Hence the cross-section of the composite system formation goes up with bombarding energy as compared to the collisions with the stable calcium projectile. Calculations also show in the quasi-fission regime, in both collisions, large number of nucleon transfer takes place from heavy target to light projectile with increasing numbers as the bombarding energy goes up towards the fusion barrier. Slightly below the fusion barrier the mean values of the proton and neutron drift toward symmetry, and can reach large values of $\Delta Z\approx 10$, $\Delta N\approx 21$ in ${}^{40}$Ca induced collisions, and $\Delta Z\approx 6$, $\Delta N\approx 9$ in ${}^{48}$Ca induced collisions, respectively. The mean-field description of the TDHF can determine only the mean values of the proton and neutron drifts. On the other hand, it is very interesting and important to provide a description of the fragment mass and charge distributions in the quasi-fission reactions. As seen from Fig. 1, sufficiently below the fusion barrier, di-nuclear structure in the collisions is maintained to a large extend.  This figure shows the density profiles in the reaction plane near maximum overlap obtained in TDHF calculations in the central collisions of ${}^{40}$Ca + ${}^{238}$U (a) and ${}^{48}$Ca + ${}^{238}$U (b) systems at energies $E_{\text{c.m.}} =202.0$ MeV and $E_{\text{c.m.}} =198.7$ MeV, respectively. Red lines indicate the position of window planes. The windows are perpendicular to the symmetry lines and pass through the minimum density planes at each instant of the collision process. Consequently, it is possible to describe nucleon exchange as a diffusion process familiar from deep-inelastic collisions  \cite{R12}. In this work, we investigate nucleon exchange mechanism in the quasi-fission reactions in head-on collisions of ${}^{40}$Ca + ${}^{238}$U and ${}^{48}$Ca + ${}^{238}$U systems at a bombarding energy slightly below the fusion barrier by employing the Stochastic Mean-Field (SMF) approach  \cite{R13}. 
The SMF approach gives rise to a Langevin description for neutron and proton exchanges characterized by diffusion and drift coefficients. We calculate these transport coefficients in the semi-classical framework in terms of the mean-field description of the TDHF solutions without any adjustable parameters. As a result of large contact times in the quasi-fission reaction, on the top of large drift toward symmetry, fragment mass and charge distributions have a very broad dispersions in both ${}^{40}$Ca + ${}^{238}$U and ${}^{48}$Ca + ${}^{238}$U systems. These results indicate that in the collisions of heavy systems, in addition to fusion, nucleon diffusion mechanism may help to populate heavy trans-uranium elements in the quasi-fission regime.

In section 2, we present a brief description of the nucleon diffusion mechanism based on the SMF approach. In section 3, we discuss transport coefficients for proton and neutron exchanges. In section 4, the result of calculations presented for central collisions of ${}^{40}$Ca + ${}^{238}$U and ${}^{48}$Ca + ${}^{238}$U systems. Conclusions are given in section 5.

\section{Diffusion mechanism }

In this work, we consider proton and neutron transfer mechanism in head-on collisions in the ${}^{40}$Ca + ${}^{238}$U and ${}^{48}$Ca + ${}^{238}$U systems at bombarding energies below the fusion barrier near the quasi-fission regime.  Specifically we carry out calculations for ${}^{40}$Ca + ${}^{238}$U and ${}^{48}$Ca + ${}^{238}$U systems at $E_{\text{c.m.}} =202.0$ MeV and $E_{\text{c.m.}} =198.7$ MeV, respectively. As seen from the TDHF calculations in \cite{R11}, near the quasi-fission regime, colliding ions stick together for a long time. At energies below the fusion barrier, as seen in Fig. 1, system maintains a binary structure to a large extent, and a visible window appears between the projectile-like and target-like partners. Therefore, we can analyze nucleon exchange mechanism, by employing nucleon diffusion concept based on the SMF approach. The phenomenological nucleon exchange model and diffusion description has been applied extensively for analyzing deep-inelastic heavy-ion collisions \cite{R12}. The SMF approach provides a more accurate microscopic framework for diffusion mechanism and extracting transport coefficients of relevant macroscopic variables without any adjustable parameters and taking the full collision geometry into account. In the SMF approach, the standard description is extended beyond the mean-field by incorporating the mean-field fluctuations in terms of generating an ensemble of events according to quantal and thermal fluctuations in the initial state (for details please refer to \cite{R13}). In extracting transport coefficients for nucleon exchange, we take the proton and neutron numbers of projectile-like fragments as independent variables. We can define the proton and neutron numbers $Z_{1}^{\lambda } (t)$, $N_{1}^{\lambda } (t)$ of the projectile-like fragments in each event by integrating over the nucleon density on the projectile side of the window as \cite{R14,R15},
\begin{equation} \label{Eq1}
\left(\begin{array}{c} {Z_{1}^{\lambda } (t)} \\ {N_{1}^{\lambda } (t)} \end{array}\right)=\int d^{3} r\theta \left[x-x_{0} \left(t\right)\right] \left(\begin{array}{c} {\rho _{p}^{\lambda } (\vec{r},t)} \\ {\rho _{n}^{\lambda } (\vec{r},t)} \end{array}\right).
\end{equation}
Here, $\lambda$ denotes the event label, $x_{0} \left(t\right)$ indicates the location of the window, and $\rho _{p}^{\lambda } (\vec{r},t)$, $\rho _{n}^{\lambda } (\vec{r},t)$ are the local densities of protons and neutrons. We take $x$-axis as the collision direction and the position $x_{0} \left(t\right)$ of the window plane is determined from the TDHF calculations. As described in \cite{R14,R15,R16,R17}, the local density is projected on the reaction plane and the window is located at the lowest density plane on the neck at each time step. 
\begin{figure}[h]
\begin{subfigure}
\centering
\includegraphics[width=8cm]{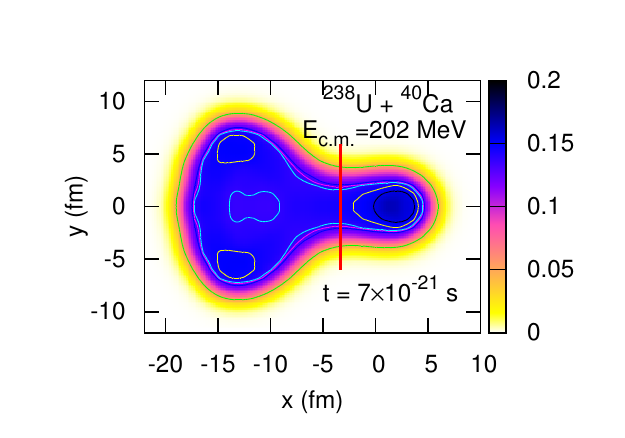}
\put(-170,45){(a)}
\vspace{-1.2cm}
\end{subfigure}
\begin{subfigure}
\centering
\includegraphics[width=8cm]{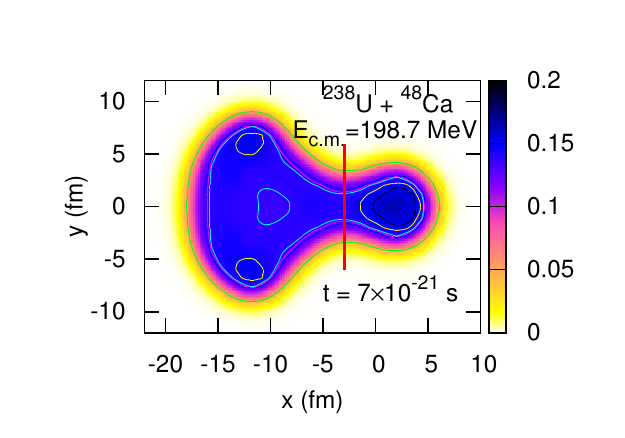}
\put(-170,45){(b)}
\end{subfigure}
\caption{(Color online) Density profiles in the reaction plane near maximum overlap in the central collisions of ${}^{40}$Ca + ${}^{238}$U (a) and ${}^{48}$Ca + ${}^{238}$U (b) systems at energies $E_{\text{c.m.}} =202.0$ MeV and $E_{\text{c.m.}} =198.7$ MeV, respectively, obtained in TDHF calculations. Red lines indicate the position of window planes.}
\end{figure}
According to the SMF approach, the proton and neutron numbers of the projectile-like fragment follows a stochastic evolution according to the following Langevin equations,
\begin{eqnarray} \label{Eq2}
\frac{d}{dt} \left(\begin{array}{c} {Z_{1}^{\lambda } (t)} \\ {N_{1}^{\lambda } (t)} \end{array}\right)&=&\int dydz\left(\begin{array}{c} {j_{x,p}^{\lambda } (\vec{r},t)} \\ {j_{x,n}^{\lambda } (\vec{r},t)} \end{array}\right) _{x=x_{0} } \nonumber \\
 &=&\left(\begin{array}{c} {v_{p}^{\lambda } (t)} \\ {v_{n}^{\lambda } (t)} \end{array}\right).
\end{eqnarray}
In this expression, the right hand side denotes the proton $v_{p}^{\lambda } (t)$ and neutron $v_{n}^{\lambda } (t)$ drift coefficients in the event $\lambda $, which are determined by the proton current $j_{x,p}^{\lambda } (\vec{r},t)$ and neutron current $j_{x,n}^{\lambda } (\vec{r},t)$ through the window.  In the SMF approach, the fluctuating proton and neutron currents are defined as \cite{R18},
\begin{eqnarray} \label{Eq3}
j_{x,a}^{\lambda } (\vec{r},t)=\frac{\hbar }{2im} \sum _{ij\in a}&&\left[\Phi _{j}^{\lambda *}  (\vec{r},t)\nabla _{x} \Phi _{i}^{\lambda } (\vec{r},t)\right.\nonumber\\
&&-\left.\Phi _{i}^{\lambda } (\vec{r},t)\nabla _{x} \Phi _{j}^{\lambda *} (\vec{r},t)\right]\rho _{ji}^{\lambda},
\end{eqnarray}
where the summations $i$ and $j$ run over a complete set of single-particle states  for protons and neutrons $a=p,n$. Drift coefficients fluctuate from event to event due to stochastic elements of the initial density matrix $\rho _{ji}^{\lambda } $ and also due to the different sets of the wave functions in different events.  As a result, there are two sources for fluctuations of the nucleon current: (i) fluctuations that arise from the state dependence of the drift coefficients, which may be approximately represented in terms of fluctuations of proton and neutron partition of the di-nuclear system, and (ii) the explicit fluctuations $\delta v_{p}^{\lambda } (t)$ and $\delta v_{n}^{\lambda } (t)$ which arise from the stochastic part of proton and neutron currents. For small amplitude fluctuations, we can linearize the drift coefficients around their mean values $v_{p} (t)$ and $v_{n} (t)$ to obtain,
\begin{eqnarray} \label{Eq4}
\left(\begin{array}{c} {v_{p}^{\lambda } (t)} \\ {v_{n}^{\lambda } (t)} \end{array}\right)&=&\left(\begin{array}{c} {v_{p} \left(t\right)} \\ {v_{n} \left(t\right)} \end{array}\right)\nonumber\\
&&+\left(\begin{array}{c} {\frac{\partial v_{p} }{\partial Z_{1} } \left(Z_{1}^{\lambda } -\overline{Z_{1}^\lambda} \right)+\frac{\partial v_{p} }{\partial N_{1} } \left(N_{1}^{\lambda } -\overline{N_{1}^\lambda} \right)} \\ {\frac{\partial v_{n} }{\partial Z_{1} } \left(Z^{\lambda } -\overline{Z_{1}^\lambda} \right)+\frac{\partial v_{n} }{\partial N_{1} } \left(N_{1}^{\lambda } -\overline{N_{1}^\lambda} \right)} \end{array}\right)\nonumber\\
&&+\left(\begin{array}{c} {\delta v_{p}^{\lambda } (t)} \\ {\delta v_{n}^{\lambda } (t)} \end{array}\right).
\end{eqnarray}
Stochastic part of the fluctuations $\delta v_{p}^{\lambda } (t)$ and $\delta v_{n}^{\lambda } (t)$ are specified by uncorrelated Gaussian distributions. These distribution have zero mean values $\overline{\delta v_{p}^{\lambda }(t)}=0$, $\overline{\delta v_{n}^{\lambda }(t)}=0$ and their variances in Markovian approximation are determined by \cite{R14,R15,R16,R17}
\begin{eqnarray} \label{Eq5}
\overline{\delta v_{p}^{\lambda } (t)\delta v_{p}^{\lambda } (t')}=2\delta (t-t')D_{ZZ} (t),
\end{eqnarray}
and
\begin{eqnarray} \label{Eq6}
\overline{\delta v_{n}^{\lambda } (t)\delta v_{n}^{\lambda } (t')}=2\delta (t-t')D_{NN} (t).
\end{eqnarray}
Here, $D_{ZZ} (t)$ and $D_{NN} (t)$ denote diffusion coefficients for proton and neutron exchange, and the mixed diffusion coefficient is zero $D_{NZ} (t)=0$.

By taking the average over the generated ensemble of the Langevin Eq. (\ref{Eq2}), the mean values evolve according to,
\begin{eqnarray} \label{Eq7}
\frac{d}{dt} \left(\begin{array}{c} {\overline{Z_{1}^\lambda} (t)} \\ {\overline{N_{1}^\lambda} (t)} \end{array}\right)&=&\int dydz\left(\begin{array}{c} {j_{x,p}^{} (\vec{r},t)} \\ {j_{x,n}^{} (\vec{r},t)} \end{array}\right) _{x=x_{0} }\nonumber\\ 
&=&\left(\begin{array}{c} {v_{p} (t)} \\ {v_{n} (t)} \end{array}\right).
\end{eqnarray}
The mean values of drift coefficients are determined by the proton and neutron fluxes. These fluxes are calculated in the mean-field description of the TDHF equations with the mean values of proton and neutron currents,
\begin{eqnarray} \label{Eq8}
j_{x,a} (\vec{r},t)=\frac{\hbar }{2im} \sum _{j\in a}&&[\Phi _{j}^{*}  (\vec{r},t)\nabla _{x} \Phi _{j} (\vec{r},t)\nonumber\\
&&-\Phi _{j} (\vec{r},t)\nabla _{x} \Phi _{j}^{*} (\vec{r},t)]n_{j},
\end{eqnarray}
where $n_{j} =1$ for occupied and $n_{j} =0$ unoccupied states.

In order to calculate the fluctuations of proton and neutron numbers, we use the stochastic part of Eq. (\ref{Eq2}) around the mean evolution,
\begin{eqnarray} \label{Eq9}
\frac{d}{dt} \left(\begin{array}{c} {\delta Z_{1}^{\lambda } (t)} \\ {\delta N_{1}^{\lambda } (t)} \end{array}\right)&=&\left(\begin{array}{c} {\frac{\partial v_{p} }{\partial Z_{1} } \left(Z_{1}^{\lambda } -\overline{Z_{1}^\lambda} \right)+\frac{\partial v_{p} }{\partial N_{1} } \left(N_{1}^{\lambda } -\overline{N_{1}^\lambda} \right)} \\ {\frac{\partial v_{n} }{\partial Z_{1} } \left(Z^{\lambda } -\overline{Z_1^\lambda}\right)+\frac{\partial v_{n} }{\partial N_{1} } \left(N_{1}^{\lambda} -\overline{N_{1}^\lambda}\right)} \end{array}\right)\nonumber\\
&&+\left(\begin{array}{c} {\delta v_{p}^{\lambda } (t)} \\ {\delta v_{n}^{\lambda } (t)} \end{array}\right),
\end{eqnarray}
where the derivatives of drift coefficients are evaluated at the mean trajectory. The variances and the co-variance of neutron and proton distribution of projectile fragments are defined as $\sigma _{NN}^{2} (t)=\overline{\left(N_{1}^{\lambda } -\overline{N_{1}^\lambda} \right)^{2} }$, $\sigma _{ZZ}^{2} (t)=\overline{\left(Z_{1}^{\lambda } -\overline{Z_{1}^\lambda}\right)^{2} }$, and $\sigma _{NZ}^{2} (t)=\overline{\left(N_{1}^{\lambda } -\overline{N_{1}^\lambda} \right)\left(Z_{1}^{\lambda} -\overline{Z_{1}^\lambda} \right)}$. Multiplying both side of Langevin equations Eq. (\ref{Eq9}) by $N_{1}^{\lambda } -\overline{N_{1}^\lambda}$ and $Z_{1}^{\lambda } -\overline{Z_{1}^\lambda}$, and taking the ensemble average, we find evolution of the co-variances are specified by the following set of coupled differential equations,
\begin{eqnarray} \label{Eq10}
\frac{\partial }{\partial t} \sigma _{NN}^{2} &=&2\frac{\partial v_{n} }{\partial N_{1} } \sigma _{NN}^{2} +2\frac{\partial v_{n} }{\partial Z_{1} } \sigma _{NZ}^{2} +2D_{NN}, \\
\label{Eq11}
\frac{\partial }{\partial t} \sigma _{ZZ}^{2} &=&2\frac{\partial v_{p} }{\partial Z_{1} } \sigma _{ZZ}^{2} +2\frac{\partial v_{p} }{\partial N_{1} } \sigma _{NZ}^{2} +2D_{ZZ}, \\
\label{Eq12}
\frac{\partial }{\partial t} \sigma _{NZ}^{2} &=&\frac{\partial v_{p} }{\partial N_{1} } \sigma _{NN}^{2} +\frac{\partial v_{n} }{\partial Z_{1} } \sigma _{ZZ}^{2} \nonumber\\
&&+\sigma _{NZ}^{2} \left(\frac{\partial v_{p} }{\partial Z_{1} } +\frac{\partial v_{n} }{\partial N_{1} } \right).
\end{eqnarray}
These co-variances describe a correlated Gaussian function for the proton and neutron distribution $P(N,Z,t)$ of the project-like or the target-like fragments,
\begin{eqnarray} \label{Eq12.1}
 P(N,Z,t)=\frac{e^{-C}}{2\pi\sigma_{NN}\sigma_{ZZ}\sqrt{1-\rho^2}},
\end{eqnarray}
where 
\begin{eqnarray} \label{Eq12.2}
 C=\frac{1}{2\left(1-\rho^2\right)}&&\left[\left(\frac{Z-\bar{Z}}{\sigma_{ZZ}}\right)^2-2\rho\left(\frac{Z-\bar{Z}}{\sigma_{ZZ}}\right)\left(\frac{N-\bar{N}}{\sigma_{NN}}\right)\right.\nonumber\\
&&\left.+\left(\frac{N-\bar{N}}{\sigma_{NN}}\right)^2\right],
\end{eqnarray}
with $\rho=\sigma^2_{NZ}/\sigma_{NN}\sigma_{ZZ}$. The mean values $\bar{N}$, $\bar{Z}$ denote the mean neutron and proton numbers of the target-like or projectile-like fragments. The set of coupled equations for co-variances are familiar from the phenomenological nucleon exchange model, and they were derived from the Fokker-Planck equation for the fragment neutron and proton distributions in the deep-inelastic heavy-ion collisions \cite{R19,R20}.

\section{Transport coefficients }

\subsection{Nucleon diffusion coefficients}
The proton and neutron diffusion coefficients $D_{ZZ}^{} $ and $D_{NN}^{} $, act as sources for determining co-variances in the coupled Eqs. (\ref{Eq10}-\ref{Eq12}). In earlier investigations, expressions these diffusion coefficients in the Markovian limit have been deduced from the SMF approach in the semi-classical framework. We present the results here, and for details we refer \cite{R14,R15,R16}. In the particular case of the head-on collisions, the expressions of proton and neutron diffusion coefficients are
\begin{eqnarray} \label{Eq13}
\left(\begin{array}{c} {D_{ZZ} } \\ {D_{NN} } \end{array}\right)&=&\int \frac{dp_{x} }{2\pi \hbar }  \left|\frac{p_{x}}{m}\right|\nonumber\\
&&\left\{\begin{array}{c} {\bar{f}_{T}^{p} \left(x_{0} ,p_{x} ,t\right)\left[1-\frac{1}{\Omega } \bar{f}_{P}^{p} \left(x_{0} ,p_{x} ,t\right)\right]} \\ {\bar{f}_{T}^{n} \left(x_{0} ,p_{x} ,t\right)\left[1-\frac{1}{\Omega } \bar{f}_{P}^{n} \left(x_{0} ,p_{x} ,t\right)\right]} \end{array}\right\}   .
\end{eqnarray}
Here, $\bar{f}_{T}^{p/n} \left(x,p_{x} ,t\right)|_{x=x_{0} } $and $\bar{f}_{P}^{p/n} \left(x,p_{x} ,t\right)|_{x=x_{0} } $ are the reduced Wigner functions in the collision direction for protons/neutrons, which are obtained by integrating coordinates and momenta over the window plane as discussed in Appendix of ref. \cite{R14}. The Wigner functions are calculated with the single-particle wave functions for protons and neutrons, which are originating from target $\left(T\right)$ and projectile $\left(P\right)$ nuclei, respectively. The quantity $\Omega$ denotes the volume of the phase space on the window plane.

\subsection{Nucleon drift coefficients }
In order to solve co-variances from Eqs. (\ref{Eq10}-\ref{Eq12}), in addition to the diffusion coefficients $D_{ZZ}^{}$ and $D_{NN}^{}$, we need to know the rate of change of drift coefficients in the vicinity of their mean values. According to the SMF approach, in order to calculate rates of the drift coefficients, we should calculate neighboring events in the vicinity of the mean-field event. Here, instead of such a detailed description, we employ the fluctuation-dissipation theorem, which provides a general relation between the diffusion and drift coefficients in the transport mechanism of the relevant collective variables as described in the phenomenological approaches \cite{R12}. Proton and neutron diffusions in the N-Z plane are driven in a correlated manner by the potential energy surface of the di-nuclear system.  As a consequence of the symmetry energy, the diffusion in direction perpendicular to the beta stability valley takes place rather rapidly leading to a fast equilibration of the charge asymmetry, and diffusion continues rather slowly along the beta-stability valley.  Fig. 2 illustrates very nicely the expected the mean-drift paths in the central collisions of ${}^{40}$Ca + ${}^{238}$U and ${}^{48}$Ca + ${}^{238}$U systems. The drift paths is obtained from the solution of the mean-field description of the TDHF equations. The di-nuclear system drifts towards symmetry during long contact time, but separates before reaching to the symmetry. Following this observation and borrowing an idea from references \cite{R20,R21}, we parameterize the $N_{1} $ and $Z_{1} $ dependence of the potential energy surface of the di-nuclear system in terms of two parabolic forms,
\begin{eqnarray} \label{Eq14}
U(N_{1} ,Z_{1} )&=&\frac{1}{2} \alpha\left(z\cos \theta -n\sin \theta \right)^{2}\nonumber\\ 
&&+\frac{1}{2} \beta\left(z\sin \theta +n\cos \theta \right)^{2}.
\end{eqnarray}
Here, $z=Z_{0} -Z_{1} $, $n=N_{0} -N_{1} $ and $\theta $ denotes the angle between beta stability valley and the $N$ - axis in the $N-Z$ plane. We can determine these angles from the mean-drift paths in Fig. 2. The quantities $N_{0} $ and $Z_{0} $ denotes the equilibrium values of the neutron and proton numbers, which are approximately determined by the average values of the neutron and proton numbers of the projectile and target ions, $N_{0} =\left(N_{P} +N_{T} \right)/2$ and $Z_{0} =\left(Z_{P} +Z_{T} \right)/2$. The first term in this expression describes a strong driving force perpendicular to the beta stability valley, while the second term describes a relative weak driving force toward symmetry along the valley. 
\begin{figure}[h]
\begin{subfigure}
\centering
\includegraphics[width=8cm]{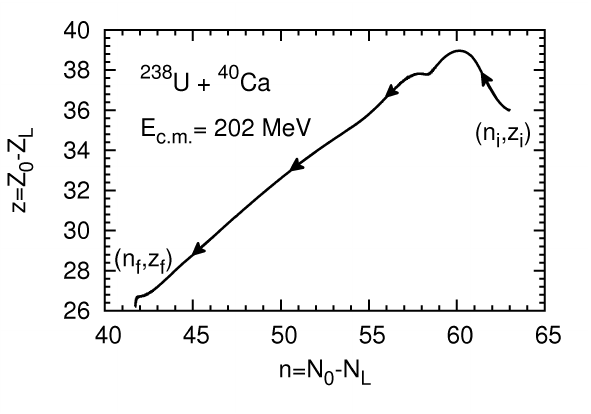}
\put(-40,50){(a)}
\vspace{-0.5cm}
\end{subfigure}
\begin{subfigure}
\centering
\includegraphics[width=8cm]{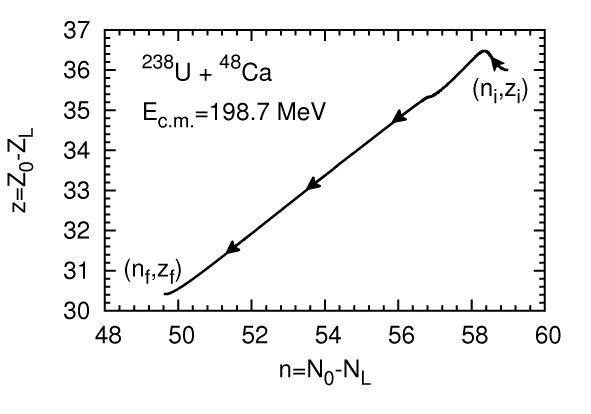}
\put(-40,50){(b)}
\end{subfigure}
\caption{Mean drift paths of projectile-like fragments in $\left(N,Z\right)$ plane in the central collision of ${}^{40}$Ca + ${}^{238}$U (a) and ${}^{48}$Ca + ${}^{238}$U (b) systems at energies $E_{\text{c.m.}} =202.0$ MeV and $E_{\text{c.m.}} =198.7$ MeV, respectively, obtained in TDHF calculations.}
\end{figure}
Following from the fluctuation-dissipation theorem, it is possible to relate the proton and neutron drift coefficients to the diffusion coefficients and the associated driving forces, in terms of the Einstein relations as follows \cite{R20,R21},
\begin{eqnarray} \label{Eq15}
v_{n}&=&-\frac{D_{NN} }{T} \frac{\partial U}{\partial N} =+\frac{D_{NN} }{T} \frac{\partial U}{\partial n} \nonumber\\
 &=& D_{NN} \left[-\alpha \sin \theta \left(z\cos \theta -n\sin \theta \right)\right.\nonumber\\
&&\qquad\quad\left.+\beta \cos \theta \left(z\sin \theta +n\cos \theta \right)\right]
\end{eqnarray}
and
\begin{eqnarray} \label{Eq16}
v_{z}&=&-\frac{D_{ZZ} }{T} \frac{\partial U}{\partial Z} =+\frac{D_{ZZ} }{T} \frac{\partial U}{\partial z}\nonumber\\
&=&D_{ZZ} \left[+\alpha \cos \theta \left(z\cos \theta -n\sin \theta \right)\right.\nonumber\\
&&\qquad\quad\left.+\beta \sin \theta \left(z\sin \theta +n\cos \theta \right)\right].
\end{eqnarray}
Here, the temperature $T$ is absorbed into coefficients $\alpha$ and $\beta$, consequently temperature does not appear as a parameter in the description. We can determine $\alpha$ and $\beta$ by matching the mean values of neutron and proton drift coefficients obtained from the TDHF solutions.  In this manner, microscopic description of the collision geometry and details of the dynamical effects are incorporated into the drift coefficients. In the liquid drop picture, the potential energy surfaces in perpendicular to the stability valley and along the stability valley have parabolic behaviors. Therefore we expect both coefficients $\alpha $ and $\beta $ to be positive. However, as a result of the quantal effects arising mainly from the shell structure, we observe that these coefficients exhibit fluctuations as a function of time, which can also be viewed as a function of the relative distance between ions. In Eqs. (\ref{Eq8}-\ref{Eq10}) for co-variances, we also need derivatives of drift coefficients with respect to proton and neutron numbers of projectile-like fragments. A great advantage of this approach, we can easily calculate these derivatives from drift coefficients to yield,
\begin{eqnarray} \label{Eq17}
\frac{\partial v_{n} }{\partial N_{1} } &=&-D_{NN} \left(\alpha \sin ^{2} \theta +\beta \cos ^{2} \theta \right),\\
\label{Eq18}
\frac{\partial v_{z} }{\partial Z_{1} } &=&-D_{ZZ} \left(\alpha \cos ^{2} \theta +\beta \sin ^{2} \theta \right),\\
\label{Eq19}
\frac{\partial v_{n} }{\partial Z_{1} } &=&-D_{NN} \left(\beta -\alpha \right)\sin \theta \cos \theta,\\
\label{Eq20}
\frac{\partial v_{z} }{\partial N_{1} } &=&-D_{ZZ} \left(\beta -\alpha \right)\sin \theta \cos \theta.
\end{eqnarray}

\section{Results}
Employing the diffusion mechanism described in the previous section, we investigate nucleon exchange mechanism in central collisions of ${}^{40}$Ca + ${}^{238}$U and ${}^{48}$Ca + ${}^{238}$U systems near the quasi-fission region at bombarding energies $E_{\text{c.m.}} =202.0$ MeV and $E_{\text{c.m.}} =198.7$ MeV, respectively. In the collision geometry, the elongation of deformed $^{238}$U target nucleus is taken as perpendicular direction to the beam (side collision). These energies are slightly below the fusion barriers. As a result, colliding ions stick together with a visible neck for a long time, and separate without forming a compound nucleus. Time dependent single-particle wave functions are determined from solutions of the TDHF equations by employing the code developed by P. Bonche et al. with the SLy4d Skyrme effective interactions \cite{R22}. Fig. 2 shows the mean-drift paths of the projectile-like fragments in $\left(N,Z\right)$ plane obtained in the TDHF calculations in the collisions of ${}^{40}$Ca + ${}^{238}$U and ${}^{48}$Ca + ${}^{238}$U systems. After a rapid charge equilibration, the system drift toward the symmetric fragmentation, which is specified by proton and neutron numbers $Z_{0} =56$ and $N_{0} =83$ for the ${}^{40}$Ca + ${}^{238}$U system and $Z_{0} =56$ and $N_{0} =87$ for the ${}^{40}$Ca + ${}^{238}$U system. The tangent of angle made by the mean-drift path with the $N$- axis is about $\tan \theta =2/3$ for both systems.  
\begin{figure}[h]
\begin{subfigure}
\centering
\includegraphics[width=8cm]{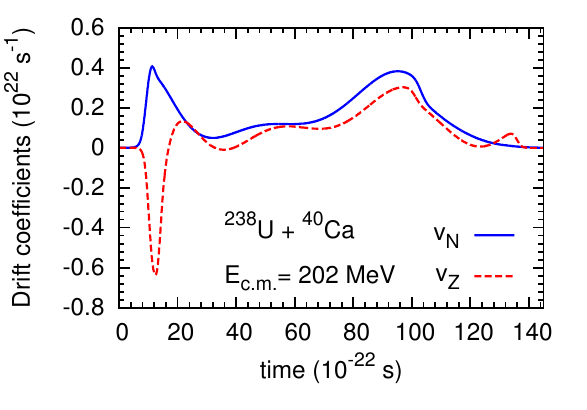}
\put(-35,135){(a)}
\vspace{-0.5cm}
\end{subfigure}
\begin{subfigure}
\centering
\includegraphics[width=8cm]{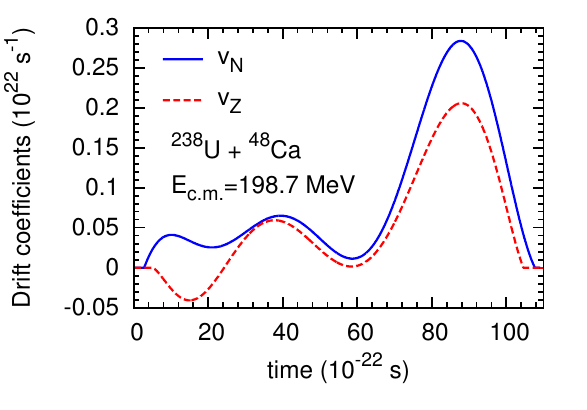}
\put(-35,135){(b)}
\end{subfigure}
\caption{(Color online) Proton drift $v_{Z} $ (dashed line) and neutron drift $v_{N} $ (solid line) coefficients in the central collisions of ${}^{40}$Ca + ${}^{238}$U (a) and ${}^{48}$Ca + ${}^{238}$U (b) systems at energies $E_{\text{c.m.}} =202.0$MeV and $E_{\text{c.m.}} =198.7$ MeV, respectively, obtained in TDHF calculations.}
\end{figure}
We use these values in parameterization of the driving potential energy in Eq. (\ref{Eq14}) for both systems. In both collisions, the system stick together approximately from an initial touching time $t_{i} =8.0\times 10^{-22}$ seconds until separation time at $t_{f} =1.4\times 10^{-20}$ seconds. We observe that during the contact time, the mean number of proton and neutron drifts are about $z(t_{f})=10$, $n(t_{f} )=21$ in the ${}^{40}$Ca + ${}^{238}$U system, and $z(t_{f} )=6$, $n(t_{f} )=9$ in the ${}^{48}$Ca + ${}^{238}$U system, respectively. Even though the sticking time is about the same in both collisions, the smaller drift in ${}^{48}$Ca induced collision is due to the smaller bombarding energy by about $3.0$ MeV. Dashed and solid lines in Fig. 3 show the proton and neutron drift coefficients in the ${}^{40}$Ca + ${}^{238}$U system (a) and in the ${}^{48}$Ca + ${}^{238}$U system (b) as a function of collision time, which are obtained by the mean-field description of the TDHF. Because of small amplitude vibrations of the window positions, drift coefficients exhibit small fluctuations in time. This figure illustrates smoothed drift coefficients obtained by averaging over short time intervals. Probably due to shell effects, in ${}^{40}$Ca induced collision protons exhibit a rapid drift toward asymmetry during the initial phase of the collision, followed by persistent drift toward symmetry in both proton and neutron numbers in both systems. We determine the dimensionless parameters $\alpha (t)$ and $\beta (t)$ in Eqs. (\ref{Eq15},\ref{Eq16}) by matching the proton and neutron drift coefficients to the results of obtained in TDHF calculations. In ${}^{40}$Ca induced collision, the parameter $\alpha (t)$ during the early times takes relatively large positive values around $\alpha (t)\approx 0.20$ until about $1.4\times 10^{-21} $ seconds while at later times it takes small fluctuating values around $\alpha (t)\approx \mp 0.05$. On the other hand parameter $\beta (t)$ take much smaller positive values $\beta (t)\approx 0.001$ as expected, also fluctuating in time. These coefficients exhibit similar behavior in the ${}^{48}$Ca induced collision. As noted above, we believe that fluctuations in these parameters are due to quantal effects arising mainly from the underlying shell structure.

\begin{figure}[h]
\begin{subfigure}
\centering
\includegraphics[width=8cm]{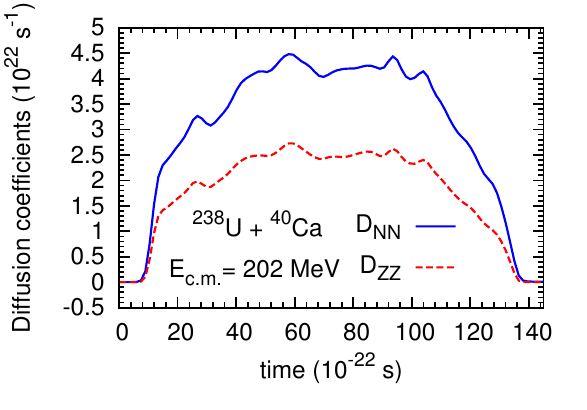}
\put(-35,135){(a)}
\vspace{-0.5cm}
\end{subfigure}
\begin{subfigure}
\centering
\includegraphics[width=8cm]{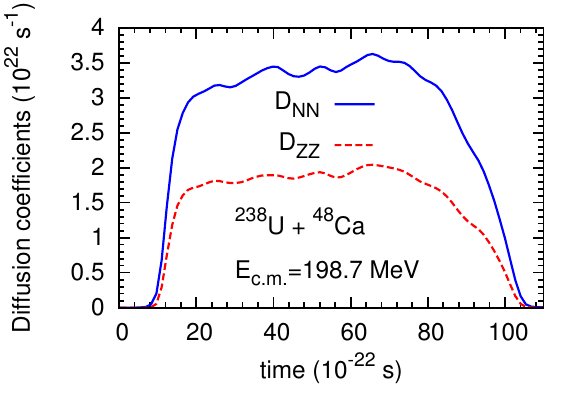}
\put(-35,135){(b)}
\end{subfigure}
\caption{(Color online) Proton diffusion $D_{ZZ} $ (dashed line) and neutron diffusion $D_{NN} $ (solid line) coefficients in the central collisions of ${}^{40}$Ca + ${}^{238}$U (a) and ${}^{48}$Ca + ${}^{238}$U (b) systems at energies $E_{\text{c.m.}} =202.0$ MeV and $E_{\text{c.m.}} =198.7$ MeV, respectively, obtained in semi-classical framework of the SMF approach.}
\end{figure}
We calculate diffusions coefficients in Eq. (\ref{Eq13}) for proton and neutron exchange in the semi-classical framework by employing the Wigner functions, which are determined in terms of the time-dependent single-particle wave functions of the TDHF solutions. The reduced Wigner functions are obtained by integrating over the phase-space volume on the window plane. The reduced Wigner functions exhibit fluctuations as a function of single-particle momentum and can take small negative values in classically forbidden regions. We eliminate these fluctuations and negative values of Wigner functions by performing a smoothing procedure as outline in \cite{R17}. Dashed and solid lines in Fig. 4 show the proton and neutron diffusion coefficients in ${}^{40}$Ca (a) and ${}^{48}$Ca (b) induced collisions as a function of time. Diffusion coefficients also exhibit small fluctuations in time due to small amplitude vibrations of the window positions. This figure illustrates smoothed diffusion coefficients obtained by averaging over short time intervals. Mainly as a result of the Coulomb barrier, the neutron diffusion coefficients are nearly twice as large as compared to the proton diffusion coefficients, in both systems. 
\begin{figure}[h]
\begin{subfigure}
\centering
\includegraphics[width=8cm]{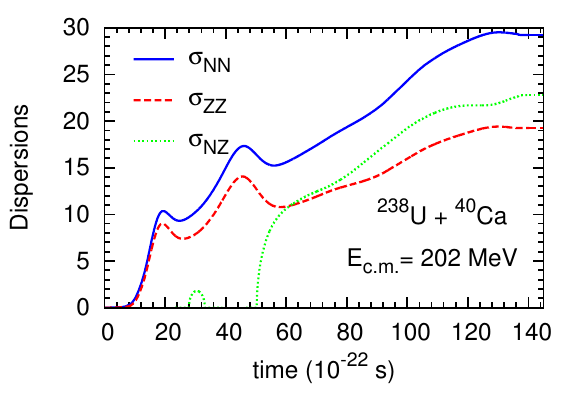}
\put(-35,135){(a)}
\vspace{-0.5cm}
\end{subfigure}
\begin{subfigure}
\centering
\includegraphics[width=8cm]{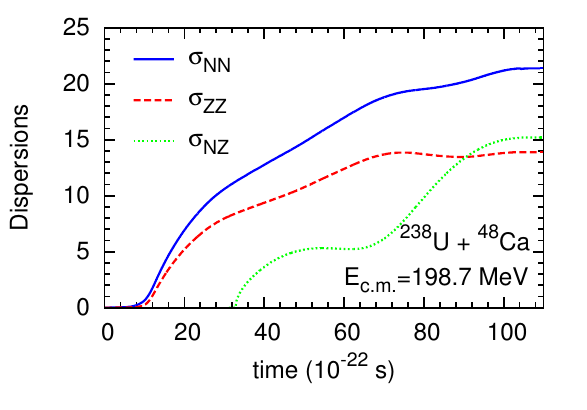}
\put(-35,135){(b)}
\end{subfigure}
\caption{(Color online) Proton dispersion $\sigma _{ZZ} $ (dashed line), neutron dispersion $\sigma _{NN}$ (solid line) and mixed dispersion $\sigma _{ZN} $ (dotted line) of the fragment distributions (projectile-like or target-like) in the central collisions of ${}^{40}$Ca + ${}^{238}$U (a) and ${}^{48}$Ca + ${}^{238}$U (b) systems at energies $E_{\text{c.m.}} =202.0$ MeV and $E_{\text{c.m.}} =198.7$ MeV, respectively.}
\end{figure}

We solve the coupled differential equations (10-12) for co-variances with the initial conditions 
$\sigma _{ZZ}^{2} (t_{i} )=\sigma _{NN}^{2} (t_{i} )=\sigma _{ZN}^{2} (t_{i} )=0$. The results are plotted in Fig. 5 as a function time.  Dashed and solid lines in the figure show proton and neutron dispersions $\sigma _{ZZ} (t)$, $\sigma _{NN} (t)$ and the square-root of co-variances $\sigma _{ZN} (t)$ in  ${}^{40}$Ca (a) and ${}^{48}$Ca (b) induced collisions, respectively. The co-variances $\sigma _{ZN}^{2} (t)$ take negative values until $5.0\times 10^{-21} $ seconds and $3.2\times 10^{-21} $ seconds in  ${}^{40}$Ca and ${}^{48}$Ca induced collisions, respectively. 

Therefore, the square-root of co-variances are not shown in these time intervals. We observe that at the separation instant in the ${}^{40}$Ca + ${}^{238}$U collision, the dispersions of proton and neutron distributions are $\sigma _{ZZ} (t_{f} )=19$ and $\sigma _{NN} (t_{f} )=29$, and the co-dispersion is 
$\sigma _{NZ} (t_{f} )=23$. The proton dispersion is nearly factor of 2.0 larger than the mean number of proton drift, while neutron dispersion is about 1.5 larger than the mean number of neutron drift. Target nucleus losses about 10 protons and 21 neutrons, but it can gain more protons and neutron by diffusion mechanism. On the other hand, at the separation instant in the ${}^{48}$Ca + ${}^{238}$U collision, the dispersions of proton and neutron distributions are $\sigma _{ZZ} (t_{f} )=14$ and $\sigma _{NN} (t_{f} )=21$, and the co-dispersion is $\sigma _{NZ} (t_{f} )=15$. The proton dispersion is nearly factor of 2.5 larger than the mean number of proton drift, while neutron dispersion is about 2.0 larger than the mean number of neutron drift. Target nucleus losses about 6 protons and 9 neutrons on the average, but it can gain more protons and neutron by diffusion mechanism. Therefore, in these collisions, diffusion mechanism near quasi-fission regime can help to populate elements heavier than uranium target nucleus. Probability distributions of the projectile-like or target-like fragments at the exit channel are determined by the correlated Gaussian of Eq. (\ref{Eq12.1}), in which the magnitudes of co-variances and mean-values are taken at the separation instant of the collision. Fig. 6 shows equal probability lines for population of target-like fragments at the exit channel in the $\left(N,Z\right)$ plane in the ${}^{40}$Ca (a) and ${}^{48}$Ca (b) induced collisions. Probability of populating a fragment with neutron and proton numbers $\left(N_{2} ,Z_{2} \right)$ relative to populating the fragment with mean neutron and proton numbers is determined by $e^{-C}$, where $C$ indicate numbers on the equal probability lines in Fig. 6. In this figure dots at the centers of ellipses indicate the elements with the mean neutron and proton numbers at the exit channel. The mean values of neutron and proton numbers at the exit channel are $\left(\overline{N}_{2}=125,\overline{Z}_{2}=82\right)$ and $\left(\overline{N}_2=137,\overline{Z}_2=86\right)$ in ${}^{40}$Ca and ${}^{48}$Ca induced collisions, respectively. As an example, we can see from this figure, the probability of populating a heavy trans-uranium element with 
$\left(N_{2}=155 ,Z_{2}=98 \right)$ relative to the populating the element with mean neutron and proton numbers is about  $e^{-0.5}=0.6$ in the $^{40}$Ca + ${}^{238}$U collision. The relative population probability of the same element in the $^{48}$Ca + ${}^{238}$U collision has about the same magnitude. Fig. 7 illustrates the dispersion $\sigma _{AA} (t)$ of total mass number distributions of the projectile-like fragments or target-like fragments as a function of time in ${}^{40}$Ca (a) and ${}^{48}$Ca (b) induced collisions, respectively.  The total dispersion is calculated from $\sigma _{AA}^{2} =\sigma _{ZZ}^{2} +\sigma _{NN}^{2} +2\sigma _{NZ}^{2} $.
\begin{figure}[h]
\begin{subfigure}
\centering
\includegraphics[width=9cm]{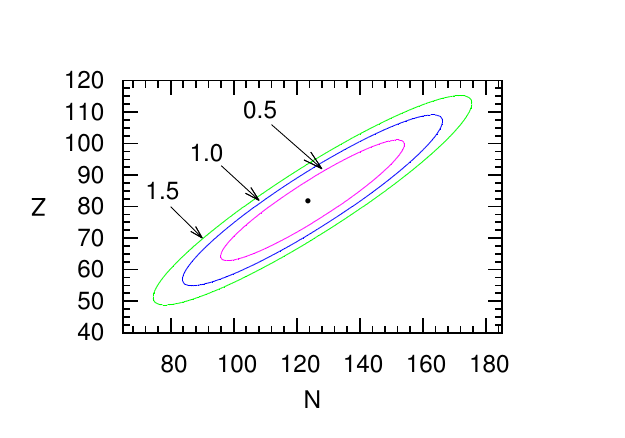}
\put(-69,51){(a)}
\vspace{-1.5cm}
\end{subfigure}
\begin{subfigure}
\centering
\includegraphics[width=9cm]{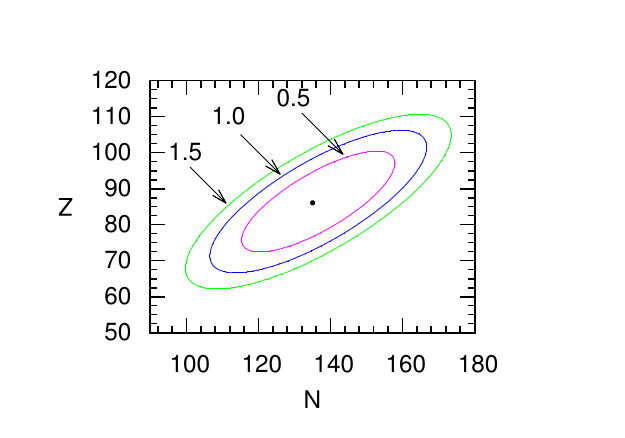}
\put(-79,51){(b)}
\end{subfigure}
\caption{(Color online) Equal probability lines for populating target-like elements with $C=0.5,1.0,1.5$ in the central collisions of ${}^{40}$Ca + ${}^{238}$U (a) and 
${}^{48}$Ca + ${}^{238}$U (b).}
\end{figure}
\begin{figure}[h]
\begin{subfigure}
\centering
\includegraphics[width=8cm]{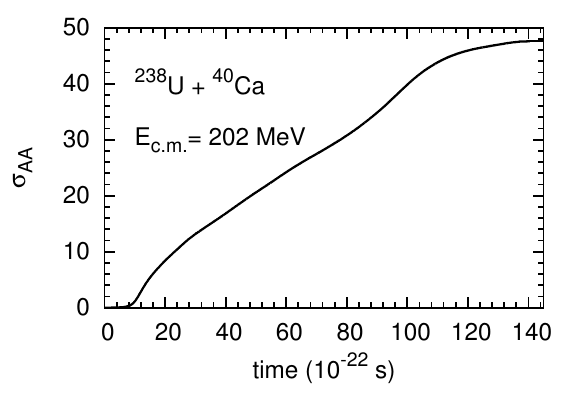}
\put(-40,50){(a)}
\vspace{-0.5cm}
\end{subfigure}
\begin{subfigure}
\centering
\includegraphics[width=8cm]{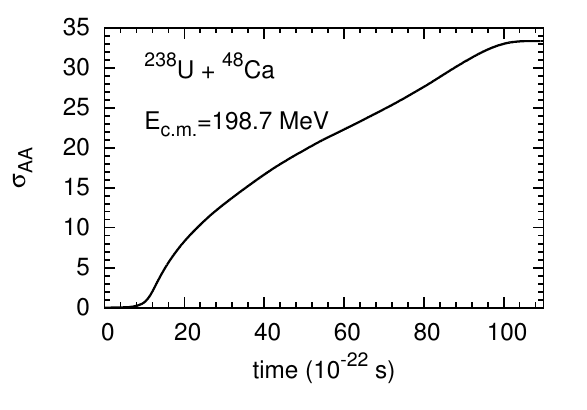}
\put(-40,50){(b)}
\end{subfigure}
\caption{Total mass dispersion  $\sigma_{AA}$  of the fragment distribution of the fragment distributions (projectile-like or target-like) in the central collisions of ${}^{40}$Ca + ${}^{238}$U  (a) and ${}^{48}$Ca + ${}^{238}$U (b) systems at energies $E_{\text{c.m.}}=202.0$  MeV and $E_{\text{c.m.}}=198.7$  MeV, respectively.}
\end{figure}

We note that the correlated Gaussian function of Eq. (\ref{Eq12.1}), which is specified by the first two moments, provides an approximate description of the fragment population. The approximation is reasonable within the range of $\pm\sigma_{AA}$ around the center points, but becomes gradually unreasonable as we move out from the center points near to the tail of the distribution functions. For example, as seen from the upper ends of $C=1.5$ lines in Fig. 6, we observe finite but small probabilities for populating fragments even  exceeding the total mass of the system. Therefore, in particular near the tail region, more accurate description of the fragment population probability is required.

In the present work, we do not discuss the energy dissipation and the excitation energy deposited in the populated fragments during nucleon diffusion process. However, we can provide an estimate of the excitation energy deposited in the fragments with mean values of protons and neutrons in the exit channel. It is possible to calculate the total excitation energy $E^*$ deposited in the mean fragments at the exit channel according to,
\begin{eqnarray}
 E^*=E_{\text{c.m.}}+Q-\text{TKE},
\end{eqnarray}
where $Q$ denotes the $Q$-value and TKE is the asymptotic value of the total kinetic energy at the exit channel. We calculate the TKE for the mean-fragment exit channel by employing the TDHF [11]. Since the interaction time is very long, the initial relative kinetic energy totally dissipates and the TKE at the exit channel is essentially determined by the Coulomb repulsion. Because of very long interaction times, it is reasonable to consider the equilibration of the excitation energy. Under this circumstance, the excitation energy between the mean fragments is shared in proportion to their mass numbers,
\begin{eqnarray}
 E_1^*=E^*\frac{\overline{A_1}}{A_{\text{tot}}},
\end{eqnarray}
and
\begin{eqnarray}
 E_2^*=E^*\frac{\overline{A_2}}{A_{\text{tot}}}.
\end{eqnarray}
In  ${}^{40}$Ca + ${}^{238}$U system, $\overline{A_1}=71$ and $\overline{A_2}=207$ with $A_{\text{tot}}=278$. In ${}^{48}$Ca + ${}^{238}$U system, $\overline{A_1}=63$ and $\overline{A_2}=223$ with $A_{\text{tot}}=286$. Calculation gives for the excitation energies $E^*_{1}=28.6$ MeV, $E^*_{2}=83.4$ MeV  in the ${}^{40}$Ca + ${}^{238}$U system, and $E^*_{1}=11.2$ MeV, $E^*_{2}=39.8$ MeV in the ${}^{48}$Ca + ${}^{238}$U system.

\section{Conclusions}
We investigate nucleon exchange mechanism in the central collisions of ${}^{40}$Ca + ${}^{238}$U and ${}^{48}$Ca + ${}^{238}$U systems below the fusion barrier near the quasi-fission regime. Sufficiently below the fusion barrier, colliding system maintains a di-nuclear structure. As a result, it is possible to describe nucleon exchange as a diffusion mechanism which is familiar from the description of deep-inelastic heavy-ion collisions. The standard mean-field description based on the TDHF equations determines the mean drift path in the N-Z plane. In order to describe fluctuations around the drift path, we employ the microscopic basis of the SMF approach, which incorporates the mean-field fluctuations beyond the average description of the standard TDHF. We calculate diffusion coefficients for proton and neutron transfer mechanisms with the help of the SMF approach in the semi-classical framework. Proton and neutron diffusion occurs in the N-Z plane in a correlated manner according to the potential energy surface of the di-nuclear system. The potential energy surface along the beta-stability line and perpendicular to the stability line are parameterized in terms of two parabolic forms. Employing Einstein relations, we deduce simple analytical expressions for proton and neutron drift coefficients. Parameters of the drift coefficients are determined with help of the mean drift path obtained in the TDHF calculations. We determine the co-variances of the neutron and proton distributions of the projectile-like fragments. Calculations show that after a fast charge equilibration, large amount of mean drift in the numbers of protons and neutrons toward symmetry. We find large dispersions of the proton and neutron distributions of the projectile-fragments during very long interaction times. The mean numbers of proton and neutron of the target nucleus decrease due to drift toward symmetry. On the other hand, large values of proton and neutron dispersions indicate that diffusion mechanism helps to populate heavy trans-uranium elements near the quasi-fission regime.

\begin{acknowledgments}
S.A. gratefully acknowledges TUBITAK and the Middle East Technical University for partial support and warm hospitality extended to him during his visits. Authors gratefully acknowledge useful discussions with D. Lacroix and A. S. Umar. This work is supported in part by US DOE Grant No. DE-FG05-89ER40530, and in part by TUBITAK Grant No. 113F061.
\end{acknowledgments}


\begin{thebibliography}{99}
\bibitem{R1} Y. T. Oganessian et al., Phys. Rev. Lett. \textbf{104}, 142502 (2010).
\bibitem{R2} Y. T. Oganessian et al., Phys. Rev. Lett. \textbf{109}, 162501 (2012).
\bibitem{R3} J. Khuyagbaatar et al., Phys. Rev. Lett. \textbf{112}, 172501 (2014).
\bibitem{R4} S. Hofmann and G. Munzenberg, Rev. Mod. Phys. \textbf{72}, 733 (2000).
\bibitem{R5} S. Hofmann et al., Eur. Phys. J. A \textbf{14}, 147 (2002).
\bibitem{R6} C.-C. Sham et al., Z. Phys. A \textbf{319}, 113 (1984).
\bibitem{R7} K.-H. Schmidt and Morawek, Rep. Prog. Phys. \textbf{54}, 949 (1991).
\bibitem{R8} G. G. Adamian, N. V. Antonenko, and W. Scheid, Phys. Rev. C \textbf{68}, 034601 (2003).
\bibitem{R9} V. Zagrebaev and W. Greiner. J. Phys. G \textbf{34}, 2265 (2007).
\bibitem{R10} Y. Aritomo, Phys. Rev. C \textbf{80}, 064604 (2009).
\bibitem{R11} V. E. Oberacker, A. S. Umar, and C. Simenel, Phys. Rev. C \textbf{90}, 054605 (2014).
\bibitem{R12} J. Randrup, Nucl. Phys. A \textbf{327}, 490 (1979).
\bibitem{R13} S. Ayik, Phys. Lett. B \textbf{658}, 174 (2008).
\bibitem{R14} S. Ayik, K. Washiyama, and D. Lacroix, Phys. Rev. C \textbf{79}, 054606 (2009).
\bibitem{R15} K.  Washiyama, S. Ayik, and  D. Lacroix, Phys. Rev. C \textbf{80}, 031602(R) (2009).
\bibitem{R16} B. Yilmaz, S. Ayik, D. Lacroix, and K. Washiyama, Phys. Rev. C \textbf{83}, 064615 (2011).
\bibitem{R17} B. Yilmaz, S. Ayik, D. Lacroix, and O. Yilmaz, Phys. Rev. C \textbf{90}, 024613 (2014).
\bibitem{R18} S. Ayik et al., Phys. Rev. C \textbf{91}, 054601 (2015).
\bibitem{R19} W. U. Schroder, J. R. Huizenga, and J. Randrup, Phys. Lett. B \textbf{98}, 355 (1981).
\bibitem{R20} A. C. Merchant and W. Norenberg, Phys. Lett. B \textbf{104}, 15 (1981).
\bibitem{R21} A. C. Merchant and W. Norenberg, Z. Phys. A \textbf{308}, 315 (1982).
\bibitem{R22} K.-H. Kim, T. Otsuka, and P. Bonche, J. Phys. G \textbf{23}, 1267 (1997).
\end{thebibliography}
\end{document}